\documentclass[12pt]{article}

\usepackage{amsmath,amssymb}
\usepackage{cancel}
\usepackage{algorithm}
\usepackage{algpseudocode}
\usepackage{bm}

\usepackage{fourier}
\usepackage[T1]{fontenc}
\usepackage[normalem]{ulem}

\usepackage{sectsty}
\sectionfont{\fontsize{12}{15}\selectfont}
\subsectionfont{\normalfont\fontsize{12}{15}\selectfont}
\subsubsectionfont{\itshape\normalfont\fontsize{12}{15}\selectfont}

\usepackage{lineno}

\usepackage[symbol]{footmisc}

\usepackage{relsize}

\usepackage{accents}
\newlength{\dhatheight}

\newcommand*\patchAmsMathEnvironmentForLineno[1]{%
	\expandafter\let\csname old#1\expandafter\endcsname\csname #1\endcsname
	\expandafter\let\csname oldend#1\expandafter\endcsname\csname end#1\endcsname
	\renewenvironment{#1}%
	{\linenomath\csname old#1\endcsname}%
	{\csname oldend#1\endcsname\endlinenomath}%
}
\newcommand*\patchBothAmsMathEnvironmentsForLineno[1]{%
	\patchAmsMathEnvironmentForLineno{#1}%
	\patchAmsMathEnvironmentForLineno{#1*}%
}
\AtBeginDocument{%
	\patchBothAmsMathEnvironmentsForLineno{equation}%
	\patchBothAmsMathEnvironmentsForLineno{align}%
	\patchBothAmsMathEnvironmentsForLineno{flalign}%
	\patchBothAmsMathEnvironmentsForLineno{alignat}%
	\patchBothAmsMathEnvironmentsForLineno{gather}%
	\patchBothAmsMathEnvironmentsForLineno{multline}%
}

\usepackage{color}
\usepackage{wrapfig}
\usepackage{indentfirst}
\usepackage{setspace}
\usepackage[numbers,super,sort&compress]{natbib}
\usepackage{booktabs}
\usepackage{rotating}
\usepackage[margin = .85in]{geometry}
\usepackage{fancyhdr}
\usepackage{tabularx}
\usepackage{pdfpages}
\usepackage{longtable}
\usepackage{multirow}
\usepackage{multicol}
\usepackage{float}
\usepackage{lscape}
\usepackage{graphicx}
\usepackage{mdwlist}
\usepackage{url}
\usepackage{comment}

\usepackage{caption}
\usepackage{subcaption}
\usepackage[compact]{titlesec}

\makeatletter
\renewcommand\@biblabel[1]{#1.}
\makeatother

\usepackage[colorlinks=true, urlcolor=blue, linkcolor=black, citecolor=black]{hyperref}

\graphicspath{{./}}

\begin{document}
\thispagestyle{empty}
\baselineskip=28pt

\noindent

\vskip .2cm
{{\noindent \huge Applied Statistics Requires Scientific Context}}


\baselineskip=12pt

\vskip .5cm
\noindent Ashley I. Naimi, PhD$^{1, *}$\\[.5em]

\vskip .5cm
\noindent $^1$ Department of Epidemiology, Emory University, Atlanta, GA.\\[.5em]

\vskip .5cm
\noindent \hskip -.2cm
\begin{tabular}{ll}
*Correspondence: & Department of Epidemiology \\[-.1cm]
& Rollins School of Public Health \\[-.1cm]
& Emory University \\[-.1cm]
& 1518 Clifton Road \\[-.1cm]
& Atlanta, GA 30322\\[-.1cm]
& \href{mailto:ashley.naimi@emory.edu}{ashley.naimi@emory.edu}
\end{tabular}

\vskip .5cm
\noindent Conflicts of Interest: None
\vskip .5cm
\noindent Acknowledgements: I am grateful to Drs. Enrique Schisterman, Sunni Mumford, Stephanie Hinkle, Alexander Levin, Eric Tchetgen Tchetgen, Sander Greenland, Atul Deodhar, and Stephen R Cole for comments, discussion, and ideas on various iterations of this work.
\vskip .5cm
\noindent Sources of Funding: None
\vskip .5cm
\vfill
\noindent \hskip -.2cm
\begin{tabular}{ll}
Text word count: 	   & 3,760 \\[-.01cm]
Abstract word count:   & 270 \\[-.01cm]
Number of Figures: 	   & 1 \\[-.01cm]
Number of Tables:      & 0  \\[-.01cm]
Number of References:  & 63 \\[-.01cm]
Running head:  & Applied Statistics Requires Scientific Context  \\ 
\end{tabular}
%
%
\newpage
\thispagestyle{empty}
\begin{center}
{\Large{\bf Abstract}}   
\end{center}
\baselineskip=12pt

\noindent  
Statistical methods are indispensable to scientific inference. However, there exists a longstanding tension across a wide range of scientific disciplines about the role that ``context'' should play in the application of statistical methods and the interpretation of statistical results. Though frequently invoked, the notion of ``scientific context'' refers to at least two distinct concepts: a set of foundational nuanced and elusive background assumptions and substantive features of a given area of study that shape the validity and reliability of statistical methods; and more quantifiable contextual issues that affect the performance of statistical methods and interpretation of statistical results. I argue here that the application and interpretation of statistical methods requires careful consideration of foundational contextual issues. To motivate the arguments, I review a recent re-formulation of the $p$-value as a measure of divergence between an observed dataset and a set of assumptions used to construct statistical measures. I use this framework to illustrate the role that context plays in two randomized trials: on low-dose aspirin for pregnancy loss, and a new inhibitor of a key biochemical pathway affecting ankylosing spondylitis. Finally, I note that the adoption of low significance thresholds in genome-wide association studies and high energy particle physics has been successful more so because of extensive validity-checking gauntlets and contextual considerations that have accompanied these low thresholds, not because of the low thresholds themselves. I use these illustrations and arguments to suggest that (i) the adoption of a universal threshold for significance testing should be abandoned as a goal of statistics reform; and (ii) the validity and optimal use of applied statistical tools requires careful consideration of nuanced scientific context.

%
%
%
\baselineskip=12pt
\par\vfill\noindent
{\bf KEY WORDS:} Statistical Inference; Statistics Reform; Scientific Context; P-Value Interpretation; Significance Testing; Significance Thresholds; Statistical Error; Randomized Controlled Trials; Statistical vs. Scientific Inference; Neo-Fisherian Inference; Neyman-Pearson Framework; Informed Judgement \\

\par\medskip\noindent
\newpage
\doublespacing
\setcounter{page}{1}

\begin{quote} 
\ldots there is no royal road to statistical induction.\cite{Cohen1990}
\end{quote}

\section*{Introduction}

There has long been a vigorous discussion and debate on the relationships between statistical methods and the generation of scientific knowledge. This discussion dates back to the inception of statistics as a field,\cite{Stigler1986} and has since focused on a number of topics. Some recent discussions have included: how $p$ values and other inferential measures are used and taught;\cite{Wasserstein2016, Rafi2020a, Alawbathani2021} the interpretation of $p$ values and related metrics such as $s$ values, and compatibility metrics;\cite{Cole2021, Gelman2019, Greenland2022a, Greenland2023, Greenland2023a} their relation to standards of evidence;\cite{Peskun2020, vanZwet2023, Gibson2021} the selection and utility of $p$ value thresholds;\cite{McShane2019, Benjamin2018, Hemerik2025, Maier2022} the impact of heuristic tools and cognitive biases in the interpretation of statistical/scientific results;\cite{Editorial2015,Rafi2020a,Pinto2023} the role of alternative frameworks, such as Bayesian decision making and inferential systems or, more recently, so called ``$e$'' values.\cite{Goodman1999a, Bickel2022, Ramdas2025}

A major theme of this discussion dating as far back as the early 20th century has focused on the role that scientific context must play when using statistical methods. However, ``scientific context'' is an imprecise term that eludes sharp definition. This has led to an important ambiguity in how this term is used in the literature, obfuscating precisely \emph{how} one might act on various calls for statistics reform.  

This paper seeks to clarify the role that \emph{scientific context} plays in statistical inference. I start by reviewing the role of scientific context in statistical inference. By way of example, and to better pin down the amorphous concept that is ``scientific context,'' I review a recently provided definition of the $p$-value as a divergence and decision metric.\cite{Greenland2023a} I use this framework and draw on examples to show precisely how scientific context plays a role in not just the interpretation, but also the validity and optimality of statistical tools. The examples drawn on here include two randomized trials on the effect of low-dose aspirin on live birth outcomes, and the effect a specific class of enzyme inhibitors on the severity of an arthritic outcome. Additionally, I discuss circumstances characterizing genome wide association studies and studies in high energy particle physics---two areas highlighted as successes of much lower significance thresholds.

I highlight details regarding the considerations that must be weighed when attempting to use statistical tools in some declarative or evidentiary capacity. Collectively, while these examples lend support to the argument that significance thresholds should, in general, be retired, more important is the recognition that ``there is no royal road to statistical induction.''\cite{Cohen1990} Scientists must consider nuanced contextual details characterizing their scientific areas. Domain specific strategies should be developed that afford a degree of flexibility in the use of informed judgement, but still maintain standards of robustness when pursuing scientific evidence.

\section*{Scientific Context and Applied Statistics}

In the context of statistics, scientific (or clinical) context is often used in one of two distinct ways. There is an extensive literature dating back to at least the early 20th century on the role that scientific context must play in the use of statistical methods. Several examples can be listed. Jerzy Neyman, in discussing the use of his null hypothesis significance testing (NHST) framework, noted that significance thresholds should be selected ``[a]ccording to the circumstances and~\ldots~subjective attitudes of the research worker.''\cite{Neyman1977}$^{(p104)}$ John Tukey famously argued that the results of a data analysis depend on judgements that arise in an attempt to mitigate the complexity of a particular experimental scenario.\cite{Tukey1962}$^{(p9,p46)}$ David Freedman documented the role that ``shoe leather''---careful detective work conducted by painstakingly walking the streets of London to collect facts and validate assumptions---played in leading John Snow to end the cholera outbreak he was studying in the 1850s.\cite{Freedman1991} Freedman argued that Snow's ``brilliant detective work on nonexperimental data'' was impressive not because of the statistical techniques used, but because of ``the handling of the scientific issues.''\cite{Freedman1991}$^{(p299)}$ In discussing the importance of models, Philip Stark notes that ``[w]hat claims and appears to be 'science' may be a mechanical amplification of the opinions and ad hoc choices~\ldots~which lacks any tested, empirical basis.'' These ``opinions and ad hoc choices'' arise as the result of and are shaped by the specific contextual details of a scientific question.\cite{Stark2022a} Several other similar such illustrations can be found throughout the scientific literature. 

In slight contrast, some researchers define scientific or clinical context as the output of some function of data. For instance, Rebecca Betensky argued that $p$-values ``must be considered in the context of certain features of the design and substantive application, such as sample size and meaningful effect size.''\cite{Betensky2019}$^{(p115)}$ Roychoudhury et al. proposed a dual-criterion phase II clinical trial design that incorporates statistical significance with ``clinical relevance'', defining the latter as no more than a ``sufficiently large effect estimate.''\cite{Roychoudhury2018}$^{(p452)}$  There is a large and growing literature on ``minimal important differences,'' which reflect the smallest magnitude estimate that would be considered ``important'' based on clinical or scientific context. Indeed, much of the literature on the flaws of using dichotomous $p$-values as a means of accruing scientific evidence point to the improved practice of considering ``contextual'' factors, such as effect magnitudes.\cite{McShane2019}

Importantly, these two notions of scientific context are connected. Features of study design and effect magnitudes often ascribed as ``context'' in the sharper sense will often arise out of aspects of the more nebulously defined notion of context. However, as I will argue, equating all of scientific context with effect magnitudes estimated from data, or some other output of a function of data, is not sufficient for conducting robust scientific inference. 

\section*{A Geometric View: The Divergence P Value}

To clarify precisely where ``scientific context'' matters for statistical methodology, consider a recently developed interpretational framework for the $p$-value.\cite{Greenland2023a, Perezgonzalez2015} In this framework, the $p$-value can be defined as a quantile location measure of divergence, distance, or separation between the data $Z$ collected for the study, and what we'd expect the data to look like if a set of conditions or assumptions were true. Denote the set of conditions and assumptions $M$. For clarity, suppose interest lies in the intention-to-treat effect of treatment $A = 1$ versus placebo $A = 0$ on an outcome of interest $Y$, where $Z = \{A,Y\}$, in a double-blind placebo controlled trial. In this case, the intention-to-treat (ITT) effect can be defined as an average treatment effect:
$$\psi = E \left (Y^{a = 1} - Y^{a = 0} \right )$$

\noindent where $Y^a$ is the outcome that would be observed if an individual was randomized to treatment arm $A = a$, and where the expectation $E(\bullet)$ is taken with respect to the population from which trial participants were sampled. This ITT effect is identified under a set of assumptions that constitute elements of $M$. In the context of a double-blind placebo controlled trial, these assumptions might include: ($M_1$) the randomization of participants to treatment and placebo groups ``worked'' (i.e., that the distribution of all measured and unmeasured covariates is balanced across both arms); ($M_2$) that blinding was maintained during all phases of the study, including at randomization as well as over the course of follow-up; ($M_3$) that any loss to follow-up accrued over the course of the trial is missing completely at random (MCAR), possibly stratified within levels of the treatment,\cite{Greenland1995b} particularly if the study protocol dictates no adjustments for missing data.

\begin{figure}[ht!]
    \centering
    \includegraphics[width=0.75\textwidth]{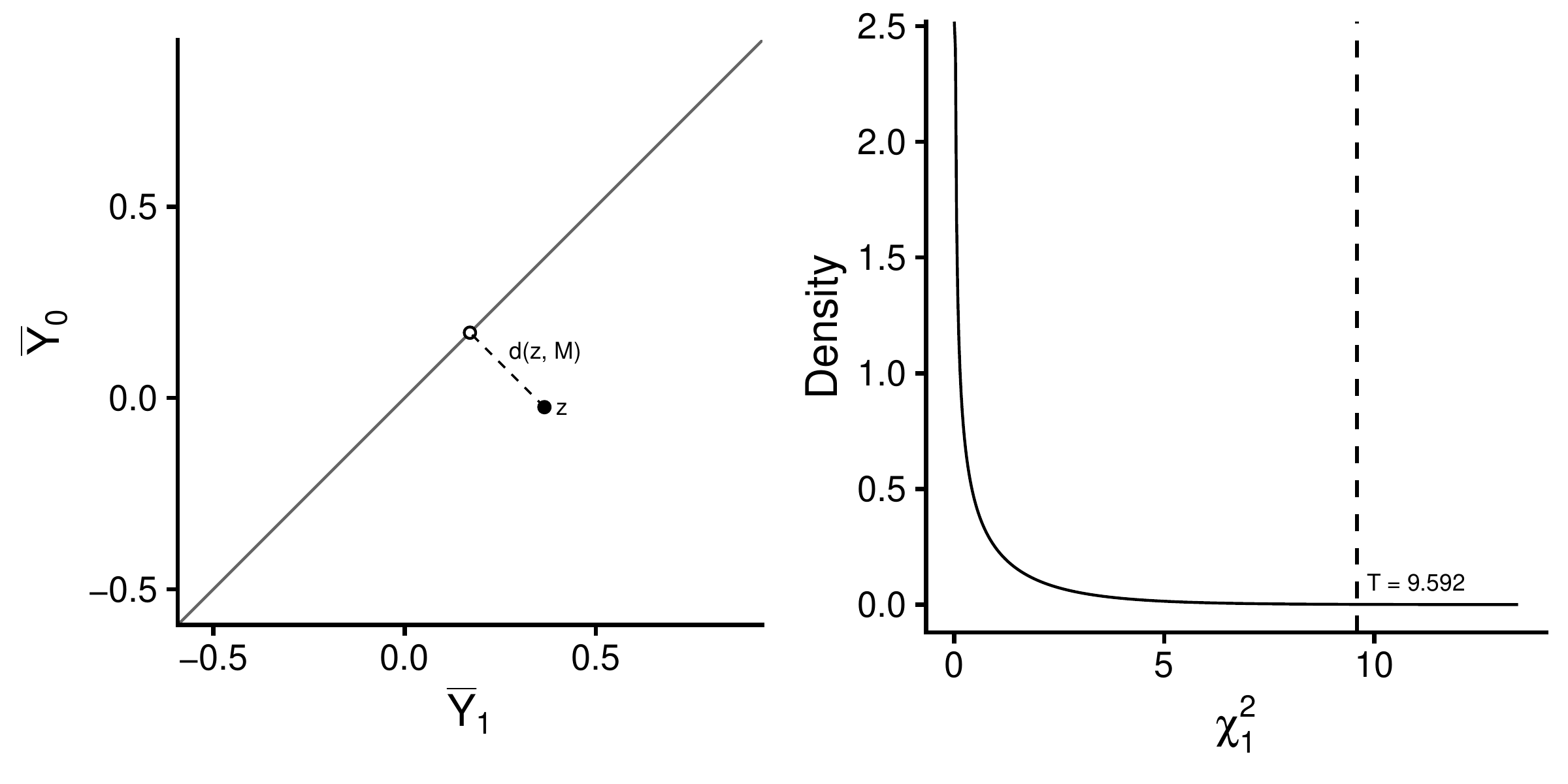}
    \caption{
        Geometric interpretation of the $p$-value. The left panel shows the observed data point $z = (\bar{Y}_1, \bar{Y}_0)$ and its orthogonal projection onto the model manifold $M$ represented by the solid diagonal line, yielding an empirical measure of discrepancy between $z$ and $M$, denoted $d(z;M)$ and indexed by the dashed line. The right panel shows the reference $\chi^2_1$ distribution of $T$, the variance standardized measure of $d(z; M)$, under $M$, with the shaded upper-tail region corresponding to the $p$-value $p = \Pr(\chi^2_1 \ge T)$. Together, these panels illustrate the $p$-value as a quantile location of the observed data-model divergence.
    }
    \label{fig:pvalue_geometry}
\end{figure}

To construct a $p$-value for this trial, let $z=(\bar Y_1,\bar Y_0)$ be the sample mean outcomes in the treatment ($A=1$) and placebo ($A=0$) arms. The complete set $M$ is defined by the conjunction of the test hypothesis $M_0: \psi = 0$ and the illustrative assumptions $\{M_1, M_2, M_3\}$. This set $M$ consists of all distributions in which the expectation of $Y$ does not depend on treatment assignment, so that $\bar Y_1 - \bar Y_0 = \psi = 0$. Figure \ref{fig:pvalue_geometry}, left panel, demonstrates how a divergence measure can be constructed in this case. Collectively, the conditions in $M$ imply that $\bar Y_1$ should be the same as $\bar Y_0$, which results in the diagonal line in the left panel of Figure \ref{fig:pvalue_geometry}. We can measure the distance between the data collected $z=(\bar Y_1,\bar Y_0)$, and this referent line using, for example a squared (Euclidean) distance metric (Figure \ref{fig:pvalue_geometry}, Left Panel), and we can account for the variability in $z$ using a variance standardized metric, leading to a test statistic for our distance measure.

Though many others could be entertained, a commonly used measure of divergence between $z$ and $M$ is the squared variance-standardized distance, or $\chi^2$ test statistic:
$$T = \frac{(\bar{Y}_1 - \bar{Y}_0)^2}{\hat{s}_1^2/n_1 + \hat{s}_0^2/n_0}$$

\noindent yielding the familiar Wald-type test statistic. The reference distribution for $T$ is also derived under the null family $M$, which in large samples approximates a $\chi^2_1$ distribution. Thus, the observed upper-tail probability
\[
p =  \Pr\!\big(\chi^2_1 \ge T\big)
\]

indexes the ``consonance'', on the unit-measure scale, between the observed data $z$ and the manifold $M$, represented by the diagonal line in Figure \ref{fig:pvalue_geometry}. Larger divergence measures yield smaller $p$-values, and suggest larger degrees of ``incompatibility'' between the set of conditions in $M$ and the observed data $z$. A $p$-value approaching values of 0 or 1 (in the limit),  represents complete incompatibility/compatibility between the data and the manifold $M$.

\section*{The Decision P Value Interpretation}

This geometric definition of the $p$-value as a quantile measure of divergence $d(z;M)$ between data realizations $z$ and model manifold $M$ makes clear how the $p$-value can be interpreted as a continuous measure of some form of evidence, referred to as the ``neo-Fisherian'' interpretation.\cite{Greenland2023a, Hurlbert2009} A second interpretation takes the $p$-value as the output of a decision criterion about whether or not to reject the model manifold $M$ as a plausible model for the data generating mechanism,\cite{Greenland2023a} referred to as the ``Neyman-Pearson'' (NP) interpretation. Conventionally, the approach proceeds with the researcher specifying a maximally acceptable false positive rejection rate $\alpha$ (Type I error), often 0.05. When in the study design phase, researchers can also conduct power calculations that determine the sample size needed to achieve a maximally acceptable false negative rejection rate $\beta$ (Type II error), often 0.2. After analyzing the data, if the $p$-value estimated from the sample is lower than the specified $\alpha$-level threshold, researchers interpret the test as having rejected the test hypothesis (e.g., null or no effect).

When the estimated $p$-value is below the pre-defined threshold, researchers are often compelled to reject the test hypothesis in isolation. Yet, this interpretation is only valid if all other elements of the model $M$ are true. Because the $p$ value measures the divergence between the observed data $z$ and the entire model manifold $M$ (not just the test hypothesis in isolation),\cite{Greenland2019, Greenland2023a} one practical dimension of the care that must accompany $p$-value interpretation is whether the elements of $M$ other than the chosen test hypothesis are true. Indeed, the diagonal equivalence line represented in the left panel of Figure \ref{fig:pvalue_geometry} is implied not just by the value of the test hypothesis (in this case, null), but also by the assumptions that randomization and blinding worked ($M_1$ and $M_2$), and that any loss to follow-up is missing completely at random or sufficiently adjusted for ($M_3$). In more realistic settings and depending on the scientific context of the study, there may be any number of additional elements included in $M$, including assumptions such as independence between units, that the sampled observations adequately represent the population of interest, or that the data collection instruments were not subject to systematic errors.

A second, and frankly secondary concern arises in the selection of the significance threshold. Though intended to be deployed ``[a]ccording to the circumstances and according to the subjective attitudes of the research worker'',\cite{Neyman1977}$^{(p104)}$ common practice is to use values of 0.05 and 0.20 for types I and II error, even when these numerical values conflict with scientific aspects of the question under study.

\section*{Examples of Scientific Context}

We consider here two specific scientific studies as case examples to clarify how specific contextual details of the study population, treatments, outcomes, and overall study circumstances must shape the way statistical tools are used. 

\subsection*{Low Dose Aspirin and Pregnancy Loss}

The Effects of Aspirin on Gestation and Reproduction (EAGeR) Trial was a multicenter double-blind placebo controlled trial of the effect of daily low-dose (81 mg) aspirin on live birth outcomes among women who were trying to conceive, but who had experienced one or two prior pregnancy losses.\cite{Schisterman2014} The trial was motivated by the fact that unexplained recurrent miscarriage may be attributable to underlying inflammation.\cite{Silver2007} Aspirin itself has been in use for over a century, it is affordable, and its side-effects are well understood and relatively low-risk. Finally, low dose aspirin had been used for nearly a decade in clinical settings to treat unexplained recurrent miscarriage, even though evidence of this effect was lacking. The EAGeR Trial was conducted to fill this evidence gap.\cite{Schisterman2013a}

This scientific context matters for study design and analysis. Indeed, clinical use of aspirin to treat unexplained recurrent miscarriage in the absence of direct evidence of an effect directly establishes a high tolerance for type I error. Power calculations for the EAGeR trial were performed for a 10\% absolute difference in the probability of live birth conducted on the basis of the standard thresholds (two-sided $\alpha = 0.05; \beta = 0.20$).\cite{Schisterman2014} However, powering the trial using a much higher type I error rate could have resulted in a smaller sample size, and significant cost savings, while still achieving the scientific goals set out by the trialists.\footnote{Per NIH Reporter, a total of roughly \$10M was spent on EAGeR. Per participant, this is about \$8.1K. Reducing sample size by 200 would have resulted in an estimated savings of roughly \$1.6M.}
 
\subsection*{Tofacitinib and Ankylosing Spondylitis}

In contrast to the EAGeR Trial, consider a phase III randomized double-blind placebo controlled trial of the effect of tofacitinib on ankylosing spondylitis (AS) among patients who have inadequately responded to or are intolerant of standard first line treatments (e.g., NSAIDs).\cite{Deodhar2021} Ankylosing spondylitis is an arthritic condition that results in inflammation and fusing of the spinal column, leading to potentially severe pain and immobility. As a therapeutic agent, tofacitinib is an inhibitor of the Janus Kinase (JAK) pathway system, which has been implicated in several autoimmune disorders closely related to AS, including rheumatoid and psoriatic arthritis. Patients in the trial were randomized 1:1 to receive 5mg tofacitinib or placebo twice daily for 16 weeks of follow-up. The primary study endpoint was based on a self reported change score, referred to as the Assessment of SpondyloArthritis international Society $\geq$20\% improvement (ASAS20) score.

Unlike aspirin, JAK inhibitors represent a much newer class of drugs, first introduced in 2011, making their long term risks much harder (impossible) to ascertain. Additionally, the known risk profile for JAK inhibitors is far more severe, and includes serious infections, cardiovascular disease, cancer, gastrointestinal perforations, anemia, and liver conditions. Power calculations for the trial were conducted for a 20\% absolute difference in ASAS20 after 16 weeks of follow-up, and suggested that 120 participants per treatment arm would yield a power of 90\% for a two-sided $\alpha$ threshold of 0.05. However, in this case, a much lower tolerance for type I error would have been warranted, given the unknown long-term effects and serious risk profile.

\subsection*{Contextualizing Threats to Validity}

Because the $p$-value measures the divergence between a realized dataset $z$ and the conjunction of assumptions that constitute elements in $M$, part of the scientific task is to understand precisely what assumptions are included in $M$ in any given setting, and to evaluate whether they hold. This is no easy task. The difficulties here echo a tension recognized since at least as far back as Plato's Socratic Dialogues (Meno's ``paradox of inquiry'').\cite{Plato2006}$^{(p230)}$ When using statistical methods, meaningful evaluation of assumptions requires prior knowledge of the system under study, yet that knowledge is precisely what scientific inquiry aims to produce. The scientist is thus caught in a reflexive loop: the validity of their tools depends on knowledge they may still be in the process of acquiring. This can make it difficult to even identify what are the relevant elements of $M$, which can lead to potential threats to a study's validity that are hiding in the messy aspects of study context. 

Because significance tests are invalidated in the presence of threats to study validity (e.g., systematic biases),\cite{Greenland2016} selective or even predominant focus on significance thresholds can be problematic, even if one is willing to adapt the threshold to a particular scientific context. These threats are present in both randomized trials and observational studies.

Considering again the trial on tofacitinib, both patients and clinicians were blinded to treatment assignment. However, tofacitinib is known to result in short-term dose-dependent changes in lipid concentrations, liver enzymes, creatine kinase, blood counts, as well as a specific side-effect profile. Results of tests for these markers, or the unique side-effect profile, could have led to functional unblinding of the study participants. Indeed, a phase II trial of tofacitinib on AS noted ``dose-dependent changes in laboratory outcomes,'' but did not report how many patients experienced these changes, nor whether these changes were reported to the participants.\cite{vanderHeijde2017} Notably, because the primary outcome was a subjective measure of self reported improvement (ASAS20/40), it is possible that unblinded participants in the treatment group reported better outcomes due to an expectancy effect.\cite{Huneke2025} Such expectancy effects 
 would threaten the validity of a study on a drug like tofacitinib. If perceived improvements in AS result from an expectancy effect of being on the drug, and not the physiologic effects of the drug itself, patients and clinicians should wonder whether a drug with a risk profile like tofacitinib's is worth it. The potential for such an expectancy effect could have been mitigated via an active comparator design.\cite{Lund2015} Indeed, standard clinical practice during the time of the trial was to first use NSAIDs, and if the patient did not respond, to transition to anti-TNF drugs, which has a similar risk profile to tofacitinib. A non-inferiority trial to compare tofacitinib to the anti-TNF regimen could have helped with some of these potential unblinding issues. 

Of relevance here is not just the choice of the significance threshold, but more importantly the elements in $M$ that constitute assumptions of the testing procedure for this phase III trial. In the case of tofacitinib, lowering the significance threshold, though scientifically warranted because of the drug's risk profile, cannot resolve the problems that arise from issues such as potential unblinding. Indeed, if expectancy effects overwhelm the physiological effects of tofacitinib, stricter testing procedures will only lead to stronger evidence for the wrong hypothesis, a so-called type III error.\cite{Stark2022a}

\subsection*{Genome Wide Association Studies and High Energy Particle Physics}

Two scientific areas have adopted lower significance thresholds with some success: genome wide association studies, as well as studies in high energy particle physics. Owing largely to multiple testing concerns that arose subsequent to the availability of HapMap data yielding insights on linkage disequilibrium, a lower significance level in genome-wide association studies of $5\times 10^{-8}$ was adopted in the mid 2000s.\cite{WellcomeTrust2007} Much earlier, in the mid 1990s high energy particle physicists published a series of papers on the top quark\cite{Abe1994, Abe1995,Abe1995b} with significance thresholds that enabled them to determine whether there was only ``evidence for'' the quark (a $p$-value of roughly 0.005), or whether they actually discovered it (a $p$-value of $5\times 10^{-7}$).\cite{Franklin2018} Later, the high energy particle physics community began to systematically employ this language, with journals \emph{Physical Review Letters} and \emph{Physical Reviews} adopting them as standards.\cite{Staley2004, Franklin2018}

Of critical note, however, is that in both GWAs and HEP, researchers devote considerable effort towards ruling out alternative explanations for a low $p$-value in a given project. Many of these alternative explanations constitute exactly those elements in $M$ that must hold in order for the results of the statistical test to remain valid. In a single GWAs study, steps can include:\cite{Uffelmann2021} consideration of confounding and selection biases during the data collection phases, quality control of the genotyping process when using microarrays or next generation sequencing methods, imputation of missing genomic data using external matched reference populations, significance testing via GWAs methods, meta-analyses by combining GWAs results with those from multiple smaller cohorts, internal or external replication, and \emph{in silico} fine mapping. After a GWAs study is conducted, researchers will sometimes follow-up on leads with alternative experimental techniques, including gene editing studies with technologies such as CRISPR-Cas9\cite{Chen2025} or massively parallel reporter assays.\cite{Lee2025} 

In high energy physics, years of preparatory work often goes into characterizing detector performance and calibrating instruments,\cite{Franklin2018, Staley2004} as well as developing extensive Monte Carlo simulations to estimate expected event distributions under an assumed test hypothesis.\cite{James2006, Cowan2011} Potential signals are distinguished from a variety of competing processes that can mimic the same experimental signatures, including background particle interactions, cosmic rays, or rare but well-understood Standard Model events.\cite{Lyons2014, Franklin2018} Once candidate signals are identified, collaborators typically perform systematic cross-checks across independent decay channels,\cite{Staley2004} compare results across different detectors or analysis pipelines,\cite{Franklin2018} and subject findings to rigorous evaluations of systematic uncertainties in measurement, reconstruction, and modeling.\cite{Cowan2011, Aad2012} In addition, blind analysis protocols and data-splitting techniques are employed to guard against confirmation bias,\cite{Lyons2014, Franklin2018} while replication across multiple groups provides an additional layer of credibility.\cite{Abe1995b, Abachi1995} Only when results are consistent across channels, methods, and experiments, and when the probability of a background fluctuation is reduced to the established discovery threshold (conventionally 5$\sigma$), do physicists present a finding as a genuine discovery.\cite{Franklin2018, Lyons2014}

The adoption of stricter significance thresholds in GWAs and HEP did not operate in the absence of scientific context considerations. Rather, they were adopted to secure robust results in fields where false positives can carry especially high scientific and economic costs. Additionally, the significance testing procedure was understood to be among the last of a series of steps in the gauntlet of technical, conceptual, scientific, and collaborative scrutiny to assure that alternative explanations for the observed findings were sufficiently considered and ruled out.

\section*{Discussion}

There is now an extensive and welcome debate on the role that $p$-values and significance testing thresholds can and should play in the process of scientific inquiry. However, largely absent from this debate is a nuanced discussion on the relationships between the mathematical and scientific considerations when using applied statistical methods. 

Several recent proposals illuminate important mathematical tradeoffs (between, e.g., type I and II errors) but consideration of these issues alone still risks evacuating the scientific substance from statistical decision-making. For example, approaches that choose $\alpha$ and $\beta$ to minimize a weighted loss can be valuable for clarifying statistical operating characteristics,\cite{Maier2022} yet they remain incomplete if the weights are detached from concrete scientific stakes. 

Indeed, the issues that can arise from elements in the model manifold $M$ are largely separate from the particular choice of applied statistical method. Nuanced scientific considerations will be relevant whether one uses $p$-values as a divergence or decision metric, or whether one opts for Bayesian methods (e.g., Bayes' factor), frequentist confidence intervals, so called causal inference methods, or a host of other techniques. Additionally, the specific considerations that might arise in one scientific context can differ from another scientific context. 

Naturally, one might ask: what are the solutions to these problems? One approach can be to develop area-specific guidelines or practices for robust scientific inference. Example areas such as genome-wide association studies or high energy particle physics can serve as a general model, and specific examples are arising in a number of biomedical sciences. For example, the development of recent ``guidelines'',	 such as CONSORT-SPIRIT,\cite{Hopewell2025, Chan2025} STRATOS,\cite{Sauerbrei2014} TARGET\cite{Cashin2025} Statements represent examples of so-called cognitive forcing tools,\cite{Pinto2023} similar to checklists popularized in surgical medicine.\cite{Gawande2010} 

However, ultimately it is important to recognize that there can be ``no mechanical alternative'' to informed judgement.\cite{Falk1995, Gigerenzer1993} Applied statistical methods must always be used as one consideration in the explicit synthesis of a range of considerations stemming from ontological, epistemological, and axiological\cite{Greenland2012} domains. These challenges underscore the reflexive difficulty of conducting cutting edge research: evaluating the assumptions on which statistical inference depends requires aspects of the very scientific knowledge that inquiry aims to produce. There is, as Cohen reminds us, no royal road, but cultivating informed, discipline specific, context-sensitive judgment can enable scientists to traverse a less-rocky path.

\newpage

\footnotesize

\bibliographystyle{aje}
\bibliography{pvalue_arxiv}

\end{document}